\documentclass[11pt]{article}
\usepackage{titlesec}
\titleformat{\paragraph}[runin]{\normalfont\itshape}{\theparagraph.}{.3em}{}[.]\titlespacing{\paragraph}{0pt}{1ex plus .1ex minus .2ex}{.5em}
\usepackage{amsmath,amssymb,bm}
\usepackage{mathabx}
\usepackage[T1]{fontenc}
\usepackage[utf8]{inputenc}
\usepackage{lmodern}
\usepackage{mathtools}
\usepackage{dsfont}
\pdfoutput=1
\usepackage[english]{babel}
\usepackage[letterpaper, hmargin=1in, top=1in, bottom=1.2in, footskip=0.6in]{geometry}
\usepackage{graphicx} 
\usepackage{booktabs} 
\usepackage{color}
\definecolor{aquamarine}{rgb}{0.5, 1.0, 0.83}
\definecolor{ao(english)}{rgb}{0.0, 0.5, 0.0}
\definecolor{armygreen}{rgb}{0.29, 0.33, 0.13}
\definecolor{awesome}{rgb}{1.0, 0.13, 0.32}
\definecolor{ballblue}{rgb}{0.13, 0.67, 0.8}
\definecolor{bittersweet}{rgb}{1.0, 0.44, 0.37}
\definecolor{blue}{rgb}{0.0, 0.0, 1.0}
\definecolor{brinkpink}{rgb}{0.98, 0.38, 0.5}
\definecolor{ballblue}{rgb}{0.13, 0.67, 0.8}
\definecolor{brightturquoise}{rgb}{0.03, 0.91, 0.87}
\definecolor{blue-green}{rgb}{0.0, 0.87, 0.87}
\definecolor{caribbeangreen}{rgb}{0.0, 0.8, 0.6}
\definecolor{cyan}{rgb}{0.0, 1.0, 1.0}
\definecolor{amber(sae/ece)}{rgb}{1.0, 0.49, 0.0}
\graphicspath{ {images/} }
\definecolor{vdarkred}{rgb}{0.6,0,0.2}

\usepackage[utf8]{inputenc}
\usepackage{lmodern}

\definecolor{vdarkred}{rgb}{0.6,0,0.2}
\definecolor{vdarkblue}{rgb}{0,0.2,0.6}
\usepackage[pdftex, colorlinks, linkcolor=vdarkblue,citecolor=vdarkred]{hyperref}

\author{Ali H.~Chamseddine$^{1,2,*}$ and J\"urg Fr\"ohlich$^{2,**}$\vspace{0.5cm}\\
$^{1}$ Physics Department, American University of Beirut, Lebanon\\
$^{2}$ Institute for Theoretical Physics, ETH Zurich, Switzerland
}
\date{June 2025}

\title{On the Dark-Energy Enigma}

\begin{document}

\maketitle

\vspace{1em}

\begin{abstract}
We present a model that offers an explanation for the presence of (Dark Matter and) Dark Energy 
in the universe. A key idea is to express the volume form of the Lorentzian metric on space-time in 
terms of a positive function of a new scalar field multiplying a certain four-form given by the wedge 
product of the differential of the mimetic scalar field and a certain closed three-form. An ansatz for this 
three-form related to one commonly used to determine the winding number of a map from a three-dimensional 
hypersurface to a three-sphere is discussed. An action functional depending on the space-time metric,
the new scalar field, the mimetic scalar and the three-form is proposed, and the field equations 
are derived. Special solutions of these equations for a Friedmann-Lema\^itre universe are presented.

\end{abstract}

\section{Introduction: Expressing the volume form on space-time in terms of
new degrees of freedom}

\label{Intro} Present-day theoretical cosmology is afflicted with some
fundamental enigmas. To mention two well known examples, it is not known what
Dark Matter (DM) and Dark Energy (DE) are made of. There are various (more or
less) plausible speculations about their nature. In this paper, we argue that
the nature of DE might be understood by viewing the metric structure and, in
particular, the volume form on space-time, $M$, as emerging from more
fundamental degrees of freedom. We focus our attention on the volume form,
\begin{equation}
\sqrt{-g}\,\,dx^{0}\wedge dx^{1}\wedge dx^{2}\wedge dx^{3}, \label{Vol-1}%
\end{equation}
where $g\,(<0)$ is the determinant of a Lorentzian metric, $g_{\mu\nu}$, on
$M$. We propose to relate this form to a 4-form that depends on some new
degrees of freedom. Our ideas are inspired in part by proposals in \cite{CM,
CCM, BF, Fr}.

To begin with, we extend the standard model of particle physics by
introducing, among some other new degrees of freedom, an additional complex
scalar field $Z=e^{(\sigma+i\theta)/f}$, where $f$ is a constant with the
dimension of mass, $\sigma$ is a scalar field that may give rise to DE
(\textquotedblleft quintessence,\textquotedblright\ \cite{Wett}), and $\theta$
is a (pseudo-) scalar field, as in \cite{BF}. A $U(1)$-symmetry acts on the
space of configurations of the field $Z$ by
\[
Z\mapsto Ze^{i\chi},\quad\theta\mapsto\theta + f\,\chi.
\]
If this symmetry is \textit{global} then the angular variable $\chi$ is an
arbitrary real number in $[0,2\pi)$, and $\theta$ is a pseudo-scalar
\textit{axion-like field} that may account for ultralight DM and, possibly,
drive the generation of primordial magnetic fields in the universe. If the
symmetry is \textit{local,} i.e., a gauge symmetry, with $\chi$ an arbitrary
smooth real-valued function on space-time, then we introduce an abelian gauge
field, $A_{\mu},$ and replace derivatives acting on $Z$ by covariant
derivatives. The role of the fields $\theta$ or $A_{\mu}$, resp., e.g., in the
generation of DM will not be discussed in the present paper, except for some
remarks in Sect.~\ref{5}. We focus our attention on a possible connection
between the field $\sigma$ and dynamical DE.

In this paper, we focus our attention on a \textit{Friedmann-Lema\^{\i}tre
universe,} with $M$ foliated by space-like three-dimensional hypersurfaces,
$\big\{\Sigma_{t}\,\big|\,t\in\mathbb{R}_{+}\big\},$ where $\Sigma_{t}%
\simeq\Sigma_{1}=:\Sigma$, and the metric on $M$ given by
\begin{equation}
\label{F-L}d\tau^{2}=dt^{2}-a^{2}(t)\,ds^{2},
\end{equation}
where $a(t)$ is the scale factor (with $a(1)=1$), and $ds^{2}$ is a Riemannian
metric on $\Sigma$. For a Friedmann-Lema\^{\i}tre universe, the hypersurface
$\Sigma$ is either a 3-sphere, or three-dimensional Euclidian space
$\mathbb{E}^{3}$, or a three-dimensional hyperbolic surface of constant
negative curvature; see, e.g., \cite{Straumann}. We focus our attention on the
choice $\Sigma=\mathbb{E}^{3}$, which appears to be the one favoured by
observational data and by inflation.

To start with, we work in a local chart, $N$, of $M$ diffeomorphic to an open
set in Minkowski space. We argue that, on $N$, the volume form \eqref{Vol-1}
is equal to a positive function, $h$, on $\mathbb{R}_{+}$ depending on the
modulus of the field $Z$ that multiplies a certain 4-form, $\alpha$, i.e.,
\begin{equation}\label{Vol-2}
\sqrt{-g}\, dx^{0}\wedge dx^{1}\wedge dx^{2}\wedge dx^{3} =
h(|Z|^{2})\, \alpha\,,
\end{equation}
where $\alpha$ is given by
\begin{equation}
\label{alpha}\alpha= d\phi\wedge\beta, \quad\text{with }\,\, \beta\, \text{ a
closed 3-form}.
\end{equation}
We propose to explore consequences of the ansatz that $\phi$ is the \textit{``mimetic
field''} introduced in \cite{CM} and $\beta$ is given by
\begin{equation}\label{B-field}
\beta= dB\,,
\end{equation}
where $B$ is a 2-form whose nature remains to be determined; (see also
Appendix A).

In the following, we investigate how a certain form of dynamical DE may result
from the ansatz in equation \eqref{Vol-2}, with $\alpha$ as in \eqref{alpha}
and \eqref{B-field}. We argue that this ansatz explains why DE is homogeneous 
and isotropic on very large distance scales.

In Sect.~\ref{3}, we introduce an action functional depending on the metric
$g_{\mu\nu}$ on $M$, the field $Z= e^{(\sigma+i \theta)/f}$ and the 4-form
$\alpha$, with one term given by the Einstein-Hilbert action. The constraint
in \eqref{Vol-2} is enforced with the help of a Lagrange multiplier field,
$\Lambda$. The field equations are derived by variation of the action
functional with respect to all the fields it depends on.

In Sect.~\ref{4}, we present solutions of the field equations, assuming that
the structure of space-time is described by a Friedmann-Lema\^{\i}tre universe.

Comments on the possible relevance of the ideas discussed in this paper for
the DM puzzle and for matter-antimatter asymmetry in the universe and some
conclusions will be presented in Sect.~\ref{5}. In Appendix A, we consider an
expression for the volume form in terms of degrees of freedom related to maps
from $\Sigma$ to a three-sphere, and some technical details not discussed in
the bulk of the paper are presented.

In this paper we only study classical field theory, neglecting effects of
quantum fluctuations.

\section{An action functional and the field equations}

\label{3}

Since visible matter only contributes around 5\% of the energy density of the
universe, we ignore the fields that make up the standard model of particle
physics, retaining only the metric $g_{\mu\nu}$, the complex scalar field $Z$,
the mimetic field $\phi$ \cite{CM} and the 3-form field $\beta$ introduced in
\eqref{B-field}. We consider the following \textit{action functional}%
\begin{align}
I  & :=-\frac{1}{2\kappa}\int d^{4}x\sqrt{-g}\,R(g_{\mu\nu})+I(Z,A_{\mu})+\int
d^{4}x\sqrt{-g}\,\lambda\left(  g^{\mu\nu} \partial_{\mu}\phi\,\partial_{\nu
}\phi-1\right)  \nonumber\\
& -2\int d^{4}x\Lambda\left(  \sqrt{-g}-\frac{1}{3!}\epsilon^{\mu\nu\kappa
\rho}h(|Z|^{2})\partial_{\mu}\phi\,\beta_{\nu\kappa\rho}\right),
\label{action}%
\end{align}
where $\kappa$ is proportional to Newton's gravitational constant, $g$ is the
determinant of the metric $g_{\mu\nu}$, $R(g_{\mu\nu})$ is the scalar
curvature, $\lambda$ and $\Lambda$ are Lagrange multiplier fields, $\phi$ is
the mimetic field \cite{CM}, and $I(Z,A_{\mu})$ is the sum of the action of
the field $Z$ possibly coupled to an abelian gauge field $A_{\mu}$ and the
Maxwell action of $A_{\mu}$. We note that the very last term on the right side
of our definition \eqref{action} of $I$ does \textit{not} depend on the metric
$g_{\mu\nu}$. The functional $I(Z,A_{\mu})$ is given by%
\begin{align}
I\left(  Z,A_{\mu}\right)    & :=\frac{f^{2}}{2}%
{\displaystyle\int}
d^{4}x\sqrt{-g}\left(  \left\vert Z\right\vert ^{-2}g^{\mu\nu}D_{\mu}%
\overline{Z}\,D_{\nu}Z-U(|Z|^{2}\right)  +\frac{1}{\gamma_{0}^{2}}\int
\sqrt{-g}\,d^{4}xF^{\mu\nu}\,F_{\mu\nu}\nonumber\\
& =\frac{1}{2}\int\sqrt{-g}\,d^{4}x\left(  \left(  g^{\mu\nu}\,\partial_{\mu
}\sigma\,\partial_{\nu}\sigma-f^{2}U(e^{2\sigma/f})\,\,+g^{\mu\nu}%
(\partial_{\mu}\theta+fA_{\mu})\,(\partial_{\nu}\theta+fA_{\nu})\right)
\right)  \nonumber\\
& +\frac{1}{\gamma_{0}^{2}}\int\sqrt{-g}\,d^{4}x\,F^{\mu\nu}\,F_{\mu\nu
}\,,\label{Z and A}%
\end{align}
with $D_{\mu}Z:=(\partial_{\mu}-iA_{\mu})Z$ the covariant derivative of the
field $Z$, and $U$ a self-interaction potential; for simplicity we henceforth set $U$ 
to 0.\footnote{It would not be difficult to include a non-trivial $U$ in our discussion, as in \cite{BF, Fr}; 
but this generalization is not of much interest in the context of the present paper.}
The constant $\gamma_{0}$ appearing in \eqref{Z and A} is the gauge coupling 
constant; if $\gamma_{0}\rightarrow0$ the gauge field disappears, and $\theta$ is 
interpreted as a pseudo-scalar \textit{axion field;} if $\gamma_{0}>0$ the field $\theta$ 
can be eliminated by a gauge transformation, and $A_{\mu}$ becomes a massive
abelian gauge field whose mass is proportional to the constant $f$; as first
studied in \cite{Stuck}. In the former case, the field $\theta$ may give
rise to ultralight axionic DM, and/or it may drive the generation of
primordial magnetic fields in the universe; in the latter case, the field
quanta of the gauge field $A_{\mu}$ may represent WIMPs, and/or $A_{\mu}$ may
appear in an explanation of the matter-antimatter asymmetry in the universe.
Henceforth we denote $h\big( \left\vert Z\right\vert ^{2}\big)$ by $h\left(\sigma\right)$,\footnote{We will 
focus our attention on the choice $h(\sigma)= \text{exp}(2\mu \sigma/f)$, where $\mu$ 
is some positive constant} and we ignore the fields $\theta$ and $A_{\mu}$; 
(but some comments on their role can be found in Sect.~\ref{5}).

The Einstein field equations are obtained by varying the action $I$ with
respect to the metric $g_{\mu\nu}$; they are given by
\begin{equation}
\kappa^{-1}\,G_{\mu\nu}=2\lambda\partial_{\mu}\phi\,\partial_{\nu}\phi+\left(
\partial_{\mu}\sigma\,\partial_{\nu}\sigma-\frac{1}{2}g_{\mu\nu}%
\,g^{\kappa\rho}\partial_{\kappa}\sigma\,\partial_{\rho}\sigma\right) %
+ 2\Lambda\,g_{\mu\nu}\,, \label{Einstein}%
\end{equation}
where we have used the constraint
\begin{equation}\label{mimetic}
g^{\mu\nu} \partial_{\mu}\phi\,\partial_{\nu}\phi=1\,,%
\end{equation}
which is obtained by varying the action $I$ with respect to $\lambda$ and
identifies $\phi$ with the mimetic scalar introduced in \cite{CM}. 
The Bianchi identity says that the covariant derivative  of the Einstein tensor $G_{\mu\nu}$ vanishes; 
furthermore, the covariant derivative of the metric  $g_{\mu\nu}$ vanishes, too. Thus, using that the 
covariant derivative of the energy-momentum tensor of $\sigma$ is equal to%
\begin{equation}
\big(\, \Box\,\sigma\big)\, \partial_{\nu}\sigma\,,
\end{equation}
where\thinspace\ $\Box\,(\cdot) := \frac{1}{\sqrt{-g}}\partial_{\mu}\big(\sqrt{-g}\,g^{\mu\nu}\partial_{\nu}(\cdot)\big)$\, 
is the covariant d'Alembertian, we conclude that
\begin{equation}\label{Bianchi-1}
2\nabla^{\mu}\left(  \lambda\partial_{\mu}\phi\,\partial_{\nu}\phi\right)
+2\big(\partial^{\mu}\Lambda\big) g_{\mu\nu}+\big(\, \Box\,\sigma \big) \partial_{\nu}\sigma=0\,.%
\end{equation}
Varying $I$ with respect to the mimetic field $\phi$ yields the equation
\begin{equation}
\lambda\,\Box\,\phi + \big(\partial_{\mu}\lambda\big) g^{\mu\nu}\partial_{\nu}\phi=- \frac{1}{\sqrt{-g}}\partial_{\mu}\left(  \Lambda\,h(\sigma)\frac{1}{3!}\epsilon^{\mu\nu
\kappa\rho}\beta_{\nu\kappa\rho}\right)  . \label{mimeticeq}%
\end{equation}
The field equation for $\sigma$, obtained by varying $I$ with respect to
$\sigma$, is given by
\begin{equation}\label{sigma}
\Box\,\sigma = \frac{1}{\sqrt{-g}}\,\frac{2\Lambda}{3!}\frac{dh}{d\sigma
}\epsilon^{\mu\nu\kappa\rho}\partial_{\mu}\phi\,\beta_{\nu\kappa\rho}\,.%
\end{equation}
If we choose $h$ to be given by $h(\sigma):=\text{exp}\big(2\mu\sigma/f)$, 
for some constant $\mu>0$, this equation becomes
\begin{equation}\label{sigma-2}
\Box\,\sigma= \frac{4\mu}{f} \Lambda\,.
\end{equation}
If $\beta=dB$ is exact then varying $I$ with respect to $B$ yields the equation\text{ }%
\begin{align}\label{B}
d\left(  \Lambda h(\sigma)d\phi\right)  =0\,,\,\,\,\, \text{i.e.,}\,\,\,
\epsilon^{\mu\nu\kappa\rho} \left(  h(\sigma) \partial_{\mu}\Lambda
+\Lambda\frac{dh}{d\sigma}\partial_{\mu}\sigma\right)  \partial_{\nu}%
\phi=0\,,\,\,\forall\,\kappa,\rho\, \,. %
\end{align}
Variation of the action functional $I$ with respect to $\Lambda$ imposes the
volume constraint \eqref{Vol-2}. In Appendix A, we consider a concrete
ansatz for the 3-form $\beta$ inspired by results in \cite{CCM}.

In order to solve the field equations \eqref{Einstein} through \eqref{B} on a
coordinate patch, $N$, of space-time, we choose the synchronous gauge
\begin{equation}
g_{00}=1,\quad g_{0j}=0,\,\,\forall\,j=1,2,3, \label{s-g}%
\end{equation}
and $-g_{ij}(x)=:\gamma_{ij}(x)$ positive-definite on $T_{x}N$, for all $x\in
N$. It follows that, in this gauge,
\begin{equation}
\phi=t+\text{ const.}\,, \label{time}%
\end{equation}
where $t$ is cosmological time; see \cite{CM}. Equation ~\eqref{Bianchi-1}, along with equations
\eqref{mimeticeq}, \eqref{sigma} then imply that
\begin{equation}\label{Bcomponents}
\frac{1}{\sqrt{\gamma}}\partial_{0}\left(  \sqrt{\gamma}\lambda\right)
+\frac{1}{h}\partial_{0}\left(  h\Lambda\right)  =0\,,\quad\quad\frac{1}%
{h}\partial_{j}\left(  h\Lambda\right)  =0\,,\,\forall\,\,j=1,2,3.%
\end{equation}
with $\gamma=\det\gamma_{ij}>0$, and we have used the volume constraint (see \eqref{Vol-2}), which reads
\begin{equation}\label{volume}
\sqrt{\gamma}=h\epsilon^{ijk}\beta_{ijk}\,. %
\end{equation}
Equation (\ref{mimeticeq}) implies that%
\begin{equation}
\frac{1}{\sqrt{\gamma}}\partial_{0}\left(  \sqrt{\gamma}\left(  \lambda
+\Lambda\right)  \right)  =\frac{1}{2\sqrt{\gamma}}\partial_{i}\left(  \Lambda
h\epsilon^{ijk}\beta_{0jk}\right)  \label{twolambdas}%
\end{equation}
Notice that equation (\ref{B}) also gives rise to the second equation in (\ref{Bcomponents}), 
which we now use in (\ref{twolambdas}) to reduce the right side to
\begin{align}\label{timeindependence}
\begin{split}
\frac{1}{2\sqrt{\gamma}}\Lambda h\epsilon^{ijk}\partial_{i}\beta_{0ij}  &
=\frac{1}{3!}\Lambda h\epsilon^{ijk}\partial_{0}\beta_{ijk} \\
&  =\frac{1}{\sqrt{\gamma}}\Lambda h\partial_{0}\left(  \frac{\sqrt{\gamma}}{h}\right)\,.
\end{split}
\end{align}
As the function $\Lambda h$ does not depend on spatial coordinates (see
\eqref{Bcomponents}) and $\Lambda$ and $h$ are independent functions, we
conclude that these functions only depend on time, and we find that
\begin{equation}
\partial_{j}\Lambda=0,\quad\partial_{j}\sigma=0\,,\,\,\,\forall\,\,j=1,2,3\,.
\label{DE}%
\end{equation}
Subtracting equation (\ref{Bcomponents}) from equation (\ref{twolambdas}),
using equation (\ref{timeindependence}), we conclude that all equations are
consistent. In the special case where the time-dependence of the volume form
of the three-dimensional metric $\gamma_{ij}$ factorizes as%
\begin{equation}
\det\gamma_{ij}=\det\widehat{\gamma}_{ij}\,\,\Omega^{2}\left(  t\right)  \,,
\end{equation}
with $\text{det}\widehat{\gamma}_{ij}$ only depending on the spatial
coordinates $x^{1},x^{2},x^{3}$ (as for a Friedmann-Lema\^{\i}tre universe),
the volume constraint in (\ref{volume}) becomes%
\begin{equation}
\frac{\Omega\left(  t\right)  }{h\left(  \sigma\left(  t\right)  \right)
}=\frac{1}{\sqrt{\widehat{\gamma}}}\epsilon^{ijk}\beta_{ijk}\,,
\end{equation}
and we observe that the left-hand side of the equation only depends on time
$t$, while the right-hand side is only space-dependent; hence both sides must
be constant. We then have that%
\begin{equation}\label{factorize}
0=\partial_{0}\left(  \ln\Omega-\ln h\right)  =\partial_{0}\left(  \ln
\sqrt{\gamma}-\ln h\right)\,.%
\end{equation}
Finally, using equation (\ref{factorize}), equations (\ref{Bcomponents})
and (\ref{twolambdas}) simplify to
\begin{equation}
\frac{1}{\sqrt{\gamma}}\partial_{0}\left(  \sqrt{\gamma}\left(  \lambda
+\Lambda\right)  \right)  =0\,,
\end{equation}
implying that
\begin{equation}
\lambda=-\Lambda+\frac{C}{\sqrt{\gamma}} \label{mimeticdark}\,,%
\end{equation}
where $C$ only depends on spatial coordinates, which allows for the presence of mimetic 
dark matter of the form $\frac{C}{\sqrt{\gamma}}.$  The fact that $\Lambda$ and $\sigma$ 
only depend on time $t$ suggests that the field $\sigma$ may generate 
\textit{dynamical Dark Energy.}

It is interesting to note that, although the field $\Lambda$ is a Lagrange
multiplier, it does contribute to the equations of motion and plays the role
of a time-dependent cosmological \textquotedblleft constant.\textquotedblright

In the next section we derive solutions of the field equations for $\sigma$,
$\Lambda$ and $g_{\mu\nu}$, specializing to a homogeneous isotropic
space-time, a Friedmann-Lema\^itre universe.

\section{Solutions of the field equations for a Friedmann-Lema\^itre\newline
universe}

\label{4}

In this section we specialize our analysis to a conformally flat
Friedmann-Lema\^{\i}tre universe. The space-time metric is then given by
\begin{equation}\label{conf-flat}
g_{00}=1,\quad g_{0i}=0, \quad g_{ij}=-a(t)^{2}\delta_{ij}\,,
\end{equation}
for $i,j=1,2,3,$ where $a(t)= \Omega(t)^{1/3}$ is the scale factor, which solves the
Friedmann-Lema\^{\i}tre equations \cite{Straumann}, and $\delta_{ij}$ is the
Kronecker delta. The volume constraint \eqref{Vol-2}, with $\alpha$ given by
\eqref{alpha} and $\beta$ as in Eq.~\eqref{B-field}, and Eq.~\eqref{time} for
$\phi$ imply that
\begin{equation}
\sqrt{-g}=a(t)^{3}=h(\sigma)\epsilon^{ijk}\partial_{i}B_{jk}%
\,,\,\,\,\,i,j,k=1,2,3\,, \label{constraint}%
\end{equation}
where $\epsilon^{ijk}\partial_{i}B_{jk}$ is a constant. If space is curved,
with $g_{ij}=-a^{2}\left(  t\right)  \widehat{\gamma}_{ij}$, the quantity
$\epsilon^{ijk}\partial_{i}B_{jk}$ is proportional to the determinant of
$\widehat{\gamma}_{ij}$.

Next, we solve the field equations, assuming that the metric on space-time is
given by \eqref{conf-flat} and that the energy-momentum tensor is diagonal,
\begin{equation}
\label{energy-momentum-1}\big(T^{\mu}_{\nu}\big) = \text{Diag}(\rho
,-p,-p,-p)\,,
\end{equation}
where $\rho$ is the energy density and $p$ the pressure of all degrees of
freedom in the universe. The Einstein equations then reduce to the Friedmann
equations (see, e.g., \cite{Straumann})
\begin{align}
\label{Friedmann}3H^{2} + 3 \frac{k}{a^{2}} = \kappa\rho+ \Lambda_{0},
\qquad2\dot{H} - 2 \frac{k}{a^{2}} = -\kappa(\rho+p)\,,
\end{align}
where the constant $\kappa$ is given by $\kappa= 8\pi G_{N}$, with $G_{N}$
Newton's gravitational constant, $H(t):=\frac{\dot{a}(t)}{a(t)}$ is the Hubble
``constant'', $k$ is proportional to the scalar curvature of $\widehat{\gamma}_{ij}$ and
vanishes for a spatially flat Friedmann-Lema\^itre universe,\footnote{Assuming
that, initially, the Universe undergoes Inflation, one expects that $k=0$.}
which is assumed henceforth, and $\Lambda_{0}$ is the cosmological constant,
which we assume vanishes.

By Eq. \eqref{Friedmann}, these assumptions are satisfied iff
\[
\rho=\rho_{\text{crit.}}:=\frac{3}{\kappa} H^{2}\,.
\]
Defining $\Omega_{0}:= \rho/\rho_{\text{crit.}},$ observational data suggest that
$\Omega_{0} \approx1$, as apparently predicted by Inflation. This implies that, 
besides Visible Matter (VM, $\approx5\%$), Dark Matter (DM, $\approx27\%$), 
there must exist Dark Energy (DE, $\approx68\%$), as confirmed by data from 
type IA supernovae (see \cite{Perl}), from the CMB and from Baryon oscillations 
(BAO) in the power spectrum of matter.

In order to solve the Friedmann equations, we must determine the equation of
state of the degrees of freedom in the universe. Since we neglect Visible
Matter and set the cosmological constant $\Lambda_{0}$ to 0, we only take into
account the contributions from the mimetic field $\phi$, the Lagrange
multiplier field $\Lambda$ and the scalar field $\sigma$. Without further
terms in the action functional $I$, the 2-form $B$ does \textit{not} contribute to the 
energy-momentum tensor. The Friedmann equations then imply that either $H$ is a
constant and $a(t)$ grows exponentially, or
\begin{equation}
H(t):=\frac{\dot{a}(t)}{a(t)}\,\propto\,t^{-1},\quad\rho(t)\,\propto
\,t^{-2}\,\,\,\,\text{ and }\,\,\,T(t)\,\propto\,\rho(t)^{1/4}\,\propto
\,\frac{1}{\sqrt{t}}\,, \label{Hubble}%
\end{equation}
where $T$ is the temperature of the CMB. Next, we determine the constributions
of the fields $\phi$ and $\sigma$ to the energy density $\rho$ and the
pressure $p$ in the universe, assuming that the field $\sigma$ only depends on
time $t$; see \eqref{DE}. Equation \eqref{Einstein} tells us that these
contributions are given by%
\[%
\begin{split}
\rho=  &  2\lambda+\rho_{0}+\frac{1}{2}\dot{\sigma}^{2} + 2\Lambda,\\
p=  &  \frac{1}{2}\dot{\sigma}^{2} - 2\Lambda\,,
\end{split}
\]
where $\rho_{0}$ accounts for additional contributions of massive
(pressureless) matter distributed homogeneously in the universe. Note that
\begin{equation}\label{rho + p}
\rho+p=2\lambda+\rho_{0}+\dot{\sigma}^{2}\,.%
\end{equation}
The quantity $\rho + p$ does not depend on the Lagrange multiplier field $\Lambda$;
(nor would it depend on the self-interaction potential $U$ that appeared in the action
functional \eqref{Z and A} and that we have set to 0).

Next, we solve the field equation \eqref{sigma} for the field $\sigma$, with
$\sigma$ only depending on cosmological time $t$; see \eqref{DE}.\footnote{Similar equations
have also been studied in \cite{BF,Fr}.} The equation then simplifies to
\begin{equation}
\label{EoM}\ddot{\sigma}+3H\dot{\sigma}=\frac{1}{a^{3}%
}\frac{2\Lambda}{f}\frac{dh}{d\sigma}\frac{1}{3!}\epsilon^{ijk}\partial
_{i}B_{jk}\,.
\end{equation}
With the ansatz that $h(|Z|^{2})=|Z|^{2\mu}=e^{2\mu\sigma/f}$, the volume
constraint \eqref{Vol-2}, with $\alpha$ given by \eqref{alpha} and
\eqref{B-field}, implies that
\begin{equation}\label{Vol-3}
\text{Right Side of Eq. }\eqref{EoM}\,=(4\mu\Lambda)/f\,,\quad\text{with
}\,\,\mu>0\,,%
\end{equation}
as already noticed in \eqref{sigma-2}. Hence
\begin{equation}\label{sigmaequation}
\ddot{\sigma}+3H\dot{\sigma} = \frac{4\mu}{f} \Lambda\,.%
\end{equation}
We consider two possible choices for $\Lambda$: first, $\Lambda=\Lambda_{0},$
is a constant; and, second, $\Lambda\, \propto\, t^{-2}, \text{ as } t\rightarrow \infty$. (We 
continue to neglect the possible presence of a self-interaction potential, $U$, but 
remark that such a potential could be included in our analysis.)

\begin{enumerate}
\item[I.] {Suppose that $\Lambda=\Lambda_{0}\not =0$. Then the right side of
(\ref{sigmaequation}) is given by a positive constant. 
To find the most general solution of equation (\ref{sigmaequation}) we have to
solve it simultaneously with the Einstein equation relating the }$0$-$0$ component of
the Einstein tensor to the $0$-$0$ component of the energy-momentum tensor,
which takes the form%
\begin{equation}
3H^{2}+3\frac{k}{a^{2}}=\frac{\kappa}{2}\overset{.}{\sigma}^{2}, \label{G00}%
\end{equation}
and, for simplicity, we set $k=0.$ This equation then implies that 
$H=\epsilon \sqrt{\frac{\kappa}{6}}\overset{.}{\sigma}$, with $\epsilon=\pm1.$ 
Equation (\ref{sigmaequation}) changes to a second order non-linear differential
equation for $\sigma$,%
\begin{equation}
\overset{..}{\sigma}+\epsilon b\overset{.}{\sigma}^{2}=c, \label{sigmaeq}%
\end{equation}
where $b=\sqrt{\frac{\kappa}{6}}$ and $c=\frac{4\mu\Lambda_{0}}{f}$, which we
assume to be positive. We will only consider the physically relevant case corresponding to
$\epsilon=1.$ We change variables, $y:=\overset{.}{\sigma}$, to get a first-order
differential equation for $y$,%
\begin{equation}
dt=\frac{dy}{c\left(  1-\frac{b}{c}y^{2}\right)  }\,. \label{yequation}%
\end{equation}
Before proceeding to find the general solution of \eqref{sigmaeq}, we consider
a special solution obtained when the denominator in \eqref{yequation}
vanishes, i.e., for $y=\pm\sqrt{\frac{c}{b}}$. By integration in $t$, we find
that%
\begin{equation}
\sigma\left(  t\right)  =\pm\sqrt{\frac{c}{b}}\, t+\sigma_{0},\quad a\left(
t\right)  =a_{0}\exp\left(  \pm\sqrt{bc}\, t\right)  \,,
\end{equation}
where the positive sign corresponds to a scale factor $a\left(  t\right)  $
increasing with time, i.e., to an expanding universe. The volume constraint
implies that $h\left( \sigma(t)\right)  $ is proportional to $a^{3}\left(  t\right)
$ and is of the form $h\left(\sigma(t) \right)  =e^{2\mu(\sigma(t)/f)}$,
and by equating the exponents we get%
\begin{equation}
\frac{2\mu}{f}\sqrt{\frac{c}{b}}=3\sqrt{bc}\,,\quad\text{i.e., }%
\,\,f=2\mu\sqrt{\frac{2}{3\kappa}}\,.
\end{equation}

We proceed to construct the general solution of equation (\ref{yequation}). By
integrating this equation in time we get
\begin{equation}
\quad y=\overset{.}{\sigma}=\zeta\coth\frac{c}{\zeta}\left(  t+t_{0}\right)
,\quad\text{with }\,\,\zeta=\sqrt{\frac{c}{b}}\,.
\end{equation}
Integrating in $t$ once more, we find that
\begin{equation}
\sigma\left(  t\right)  =\sigma_{0}+\frac{1}{b}\ln\sinh\sqrt{bc}\left(
t+t_{0}\right)  ,
\end{equation}
and, for\thinspace\ $\epsilon=1$, the scale factor $a\left(  t\right)  $ is
given by%
\begin{equation}
a\left(  t\right)  =a_{0}\left(  \sinh\sqrt{bc}\left(  t+t_{0}\right)
\right)  \,.
\end{equation}
Recalling the volume constaint, this implies that $h\left(  \sigma\right)
\sim a^{3}\left(  t\right)  $ is of the form $e^{\frac{2\mu}{f}\sigma\left(
t\right)  }.$ For small times, $t$, we then have that $\sigma\left(  t\right)
=\frac{1}{b}\ln\sqrt{bc}\left(  t+t_{0}\right)  $ and $a\left(  t\right)
=a_{0}^{\prime}\left(  t+t_{0}\right)  $, while, for large $t$, we obtain that
$\sigma\left(  t\right)  =\sqrt{\frac{c}{b}}\, t+\sigma_{0}$ and $a\left(
t\right)  =a_{0}^{\prime}\exp\left(  \sqrt{bc}\left(  t+t_{0}\right)  \right)
$, coinciding with the special solution found at the beginning.

\item[II.] {Next, we assume that $\Lambda$ may depend on time $t$, with
$\Lambda(t)\,\propto\,t^{-2}$, as $t\rightarrow\infty$. Asymptotically,
as $t\rightarrow\infty$, the scale factor $a(t)$ then grows like a power of $t$,
and hence $H(t)\sim t^{-1}$. For simplicity we assume that $\Lambda=L\, t^{-2}$},
for a positive constant $L$. We are interested in constructing the general solution 
of equation (\ref{sigmaequation}); luckily one can find the exact solution. We change 
variables,%
\begin{equation}\label{C-of-V}
u=\frac{1}{t\overset{.}{\sigma}}\,,%
\end{equation}
and \eqref{sigmaequation} simplifies to%
\begin{equation}
\frac{dt}{t}=-\frac{1}{c}\frac{du}{\left(  u-u_{+}\right)  \left(
u-u_{-}\right)  } \label{uequation}%
\end{equation}
where%
\begin{equation}
\quad u_{\pm}=\frac{1}{2c}\left(  -1\pm\sqrt{1+4\epsilon bc}\right)  ,\quad
b=\sqrt{\frac{\kappa}{6}},\quad c=\left\vert \frac{4\mu L}{f}\right\vert
\end{equation}
For simplicity, we choose $\epsilon$ to be unity. We begin by constructing special
solutions for $u=u_{\pm},$ implying that $\overset{.}{\sigma}=\frac{1}{u_{\pm}}\frac{1}{t}$, 
which integrates to
\begin{equation}
\sigma\left(  t\right)  =\frac{1}{u_{\pm}}\ln\frac{t}{t_{0}}+\sigma_{0},\quad
a\left(  t\right)  =a_{0}\left(  \frac{t}{t_{0}}\right)  ^{\frac{b}{u_{\pm}}}%
\end{equation}
Comparing the powers of $t$ in the volume constraint coming from $h\left(\sigma\right)$ 
and from $a^{3}\left(  t\right)  $, we find that%
\begin{equation}
\frac{2\mu}{f}=3b\quad \Rightarrow \quad f=2\mu\sqrt{\frac{2}{3\kappa}}%
\quad \Rightarrow \quad c=6bL\,.
\end{equation}
To end up with positive powers of time $t$ in $a\left(  t\right)$ (as appropriate for an expanding
universe), we only keep $u_{+}$\thinspace, which entails that the power becomes $\frac{2bc}{\sqrt{1+4bc}\, -1}$,
which, for a small constant $L$, is approximately equal to $1,$ showing that
the energy density of the universe is due to the kinetic term of the field
$\sigma.$

For large values of the constant $L$, the power is $\sqrt{bc}=\sqrt{\kappa L}%
$. Assuming that $u\neq u_{+},$ $u_{-}$ (i.e., we do not consider the special
solutions discussed above), we integrate equation (\ref{uequation}) to get%
\begin{equation}
\frac{u-u_{+}}{u-u_{-}}=\left(  \frac{t}{t_{0}}\right)  ^{\sqrt{1+4bc}}\,,
\end{equation}
which can be solved to yield
\begin{equation}
u=\frac{1}{t\overset{.}{\sigma}}=\frac{\left(  \frac{t}{t_{0}}\right)
^{\sqrt{1+4bc}}u_{-}-u_{+}}{\left(  \frac{t}{t_{0}}\right)  ^{\sqrt{1+4bc}}%
-1}\,.
\end{equation}
This equation can be integrated, and we find that
\begin{equation}
\sigma\left(  t\right)  =-\frac{1}{b}\left(  \ln\cosh\left(  \ln\left(
\frac{t}{t_{0}}\right)  ^{\alpha}+\tanh^{-1}\frac{1}{\alpha}\right)
-\ln\left(  \frac{t}{t_{0}}\right)  \right)  \,,
\end{equation}
where $\alpha=\frac{1}{2}\sqrt{1+4bc}.$ Using the formula for $\overset{.}{\sigma}$ obtained above 
we obtain the following expression for the scale factor $a(t)$%
\begin{equation}
a\left(  t\right)  =a_{0}\frac{\left(  \frac{t}{t_{0}}\right)  }{\cosh\left(
\ln\left(  \frac{t}{t_{0}}\right)  ^{\alpha}+\tanh^{-1}\frac{1}{\alpha
}\right)  }\,.
\end{equation}
For $h\left(  \sigma\right)  $ to satisfy the volume constraint, it must be of the form 
$h\left(  \sigma\right)  =e^{\frac{2\mu}{f}\sigma}.$ (The fact that a reasonable value for 
the Hubble constant is obtained if one assumes that $f^{2}=\mathcal{O}(\kappa^{-1})$ suggests 
that the field $\sigma$ is a gravitational degree of freedom. In \cite{CFG}, a gravitational 
degree of freedom with the properties of $\sigma$ has been obtained as a result of the
assumption that space-time is two-sheeted.)
\end{enumerate}

\section{Summary and conclusions}

\label{5} In this paper we have studied implications of the idea that the
volume form on space-time can be expressed in terms of possibly more
fundamental degrees of freedom, namely a complex scalar field $Z$, the mimetic
scalar $\phi$ and a 2-form field $B$. Our results suggest that this idea could
lead to a viable starting point for a theory of Dark Energy. Among our predictions are
a relation between Newton's gravitational constant $\kappa$ and the ``decay constant'' 
$f$ of the scalar fields $\sigma$ and $\theta$, namey $f\, \propto\, \kappa^{-1/2}$. If
the field $Z$ is coupled to a gauge field $A_{\mu}$, as in \eqref{Z and A}, then
this gauge field would have a bare mass of the order of the \textit{Planck mass.} 

All results presented in this paper suggest that the field $Z$ is a gravitational degree of
freedom; see also \cite{CFG}.

In this paper we have not studied effects caused by the axion-like field $\theta$ 
or the gauge field $A_{\mu}$, respectively; see \eqref{action} and \eqref{Z and A}. 
Possible effects caused by an axion-like field $\theta$ (gauge coupling $\gamma_0 = 0$,
see \eqref{Z and A}) related to the DM puzzle and/or to the presence of highly homogeneous intergalactic 
magnetic fields in the universe are discussed in \cite{BF, Fr} and \cite{TW, JS, FP, Durrer},
respectively, and in references given therein. 

A very heavy $U(1)$-gauge field of the kind we have considered in \eqref{Z and A}, coupled
to the conserved $B-L$ current may be an ingredient in mechanisms explaining the matter-antimatter
asymmetry in the universe; see \cite{Weinberg}.

Perhaps, the most pressing and interesting open problem might be to reconstruct the entire 
space-time metric $g_{\mu\nu}$ from more fundamental degrees of freedom, including those described 
by the fields $Z$, $\phi$ and $B_{\mu\nu}$; see also \cite{CCM}. We hope to return to this and 
other related problems in future work.\\

\textit{Acknowledgements.} The work of A.~H.~C is supported in part by the
National Science Foundation Grant No. Phys-2207663. A.~H.~C.~thanks
V.~F.~Mukhanov for numerous discussions on problems related to the ones
studied in this paper. J.~F.~thanks H.~Bernardo, R.~Brandenberger and
B.~Pedrini for discussions on ideas somewhat rekated to ones in this paper.

\section*{Appendix A}

\label{A} In this appendix we present an alternative way of relating the field $\sigma$ appearing in the definition 
of the complex scalar field $Z$ to the scale factor of the space-time metric $g_{\mu\nu}$. This is not an easy task, 
because such relations tend to break diffeomorphism invariance, as in so-called Unimodular Gravity (see
\cite{Einstein, Pauli, H-T}). One of the consequences of the spectral formulation of the Standard Model is 
that the volume form of a four-dimensional Euclidean space-(imaginary-)time manifold $M$ is given by 
the sum of two specific four-forms; diffeomorphism invariance remains unbroken. 
These forms can be expressed in terms of two mappings of $M$ to two four-spheres. 

Here we propose to use a similar idea in our context of pseudo-Riemannian (Lorentzian) space-time manifolds.
We limit our analysis to space-times, $M$, that are foliated by three-dimensional space-like hypersurfaces, 
$\big\{\Sigma_t\big\}_{t\in \mathbb{R}}$. At every time $t$, we consider two different mappings, $Y$ and 
$\widetilde{Y}$, from the hypersurfaces $\Sigma_t$ to the three-sphere $S^{3}$. The closed 3-form $\beta$ 
introduced in \eqref{alpha} then takes the form
\begin{equation}\label{beta}
\beta=\epsilon_{abcd}\,\big[Y^{a}dY^{b}\wedge dY^{c}\wedge dY^{d}%
\,+\,\widetilde{Y}^{a}d\widetilde{Y}^{b}\wedge d\widetilde{Y}^{c}\wedge
d\widetilde{Y}^{d}\,\big]\,,
\end{equation}
where $a,b,c,d$ range over $1,2,3,4$, $Y$ and $\widetilde{Y}$ are unit vectors in $\mathbb{R}^{4}$, 
and the volume constraint is encoded in the equation%
\begin{equation}
\sqrt{-g}=h\left(  \sigma\right)  \frac{1}{3!}\epsilon^{\mu\nu\kappa\lambda
}\epsilon_{abcd}\partial_{\mu}X\left(  Y^{a}\partial_{\nu}Y^{b}\partial
_{\kappa}Y^{c}\partial_{\lambda}Y^{d}+\,\widetilde{Y}^{a}d\widetilde{Y}%
^{b}\wedge d\widetilde{Y}^{c}\wedge d\widetilde{Y}^{d}\right)\,,
\end{equation}
with
\begin{equation}
\sum_{a=1}^{4}Y^{a}Y^{a}=1,\qquad \sum_{a=1}^{4}\widetilde{Y}^{a}\widetilde{Y}^{a}=1\,,
\end{equation}
and
\begin{equation}
\frac{1}{3!}\epsilon_{abcd}%
{\displaystyle\int\limits_{\Sigma_t}}
Y^{a}dY^{b}dY^{c}dY^{d}%
\end{equation}
is the winding number of the map $Y$ restricted to the three-dimensional hypersurface $\Sigma_t$. 
(Two mappings, $Y$ and $\widetilde{Y}$, are needed to ensure that all hypersurface geometries are 
compatible with the volume constraint.) The fields $Y$ and $\widetilde{Y}$ can be viewed as fields 
of a non-linear $\sigma$-model with target $S^{3}\times S^{3}$. The ansatz described in 
Eqs.~\eqref{alpha} and \eqref{beta} is motivated by general results in \cite{CCM}, which are not
reproduced here.The best option, although not the only one, is to choose $X$ to be given by the mimetic 
scalar, i.e., $X=\phi$, so that
\begin{equation}
g^{\mu\nu}\partial_{\mu}\phi\partial_{\nu}\phi=1
\end{equation}
(This choice does not increase the number of degrees of freedom, as this constraint
replaces the longitudinal mode in $g_{\mu\nu}$ with $\partial_{\mu}\phi.$) In what follows, 
we simplify our analysis by considering only one mapping $Y$ (with the understanding that 
when the winding number of one mapping is singular the one of the second mapping would
not be singular).

We consider the following action%
\begin{align}
I  &  =-\frac{1}{2}%
{\displaystyle\int}
d^{4}x\sqrt{-g}R\left(  g\right)  +%
{\displaystyle\int}
d^{4}x\sqrt{-g}\left(  \frac{1}{2}g^{\mu\nu}\partial_{\mu}\sigma\partial_{\nu
}\sigma-U\left(  \sigma\right)  \right) \nonumber\\
&  +%
{\displaystyle\int}
d^{4}x\sqrt{-g}\lambda\left(  g^{\mu\nu}\partial_{\mu}\phi\partial_{\nu}%
\phi-1\right)  -2%
{\displaystyle\int}
d^{4}x\Lambda\left(  \sqrt{-g}-\frac{1}{3!}\epsilon^{\mu\nu\kappa\lambda
}h\left(  \sigma\right)  \partial_{\mu}\phi\epsilon_{abcd}Y^{a}\partial_{\nu
}Y^{b}\partial_{\kappa}Y^{c}\partial_{\lambda}Y^{d}\right) \nonumber\\
&  +\frac{1}{2}%
{\displaystyle\int}
d^{4}x\sqrt{-g}\lambda^{\prime}\left(  Y^{a}Y^{a}-1\right)
\end{align}
where $Y^{a}Y^{a}$ is an abbreviation for $\sum_{a} Y^{a}Y^{a}$, and, in this appendix, we set the 
Planck length $\kappa=1.$ The Einstein equations obtained by varying the metric $g^{\mu\nu}$ 
are given by
\begin{align}
G_{\mu\nu}  &  =2\lambda\partial_{\mu}\phi\partial_{\nu}\phi\nonumber\\
&  +\left(  \partial_{\mu}\sigma\partial_{\nu}\sigma-\frac{1}{2}g_{\mu\nu
}g^{\alpha\beta}\partial_{\alpha}\sigma\partial_{\beta}\sigma\right)
+g_{\mu\nu}\left(  2\Lambda+U\left(  \sigma\right)  \right)
\end{align}
Varying with respect to the mimetic field $\phi$ yields%
\begin{equation}
\partial_{\mu}\left(  2\lambda\sqrt{-g}g^{\mu\nu}\partial_{\nu}\phi\right)
=-2\partial_{\mu}\left(  \Lambda h\left(  \sigma\right)  \frac{1}{3!}%
\epsilon^{\mu\nu\kappa\lambda}\epsilon_{abcd}Y^{a}\partial_{\nu}Y^{b}%
\partial_{\kappa}Y^{c}\partial_{\lambda}Y^{d}\right)  , \label{phi}%
\end{equation}
Varying with respect to $\sigma$ gives%
\begin{equation}
\partial_{\mu}\left(  \sqrt{-g}g^{\mu\nu}\partial_{\nu}\sigma\right)
=2\Lambda\frac{1}{3!}\epsilon^{\mu\nu\kappa\lambda}h^{\prime}\left(
\sigma\right)  \partial_{\mu}\phi\epsilon_{abcd}Y^{a}\partial_{\nu}%
Y^{b}\partial_{\kappa}Y^{c}\partial_{\lambda}Y^{d}-U^{\prime}\left(
\sigma\right)  \label{rho}%
\end{equation}
Finally, setting the variation of the action with respect to the fields $Y^{a}$ to 0 implies that %
\begin{align}
&  \sqrt{-g}\lambda^{\prime}Y_{a}+2\Lambda h\left(  \sigma\right)  \frac
{1}{3!}\epsilon^{\mu\nu\kappa\lambda}\epsilon_{abcd}\partial_{\mu}\phi
\partial_{\nu}Y^{b}\partial_{\kappa}Y^{c}\partial_{\lambda}Y^{d}\nonumber\\
&  =-3\partial_{\mu}\phi\partial_{\nu}\left(  \Lambda h\left(  \sigma\right)
\frac{1}{3!}\epsilon^{\mu\nu\kappa\lambda}\epsilon_{abcd}Y^{b}\partial
_{\kappa}Y^{c}\partial_{\lambda}Y^{d}\right)  \label{Y}%
\end{align}
Multiplying this equation with $Y^{a}$ we find that
\begin{equation}
\sqrt{-g}\lambda^{\prime}=-8\Lambda\sqrt{-g}\,,%
\end{equation}
or, equivalently,
\begin{equation}
\lambda^{\prime}=-8\Lambda\,.
\end{equation}
The volume constraint is expressed by%
\begin{equation}
\sqrt{-g}=\frac{1}{3!}\epsilon^{\mu\nu\kappa\lambda}\epsilon_{abcd}h\left(
\sigma\right)  \partial_{\mu}\phi Y^{a}\partial_{\nu}Y^{b}\partial_{\kappa
}Y^{c}\partial_{\lambda}Y^{d}\,.%
\end{equation}
In the synchronous gauge we have that%
\begin{equation}
g_{00}=1,\quad g_{0i}=0,\quad g_{ij}=-\gamma_{ij}\,,%
\end{equation}
and hence
\begin{equation}
\phi=t+\mathrm{const.}%
\end{equation}
Specializing to the Friedmann metric,
\begin{equation}
\gamma_{ij}=\widehat{\gamma}_{ij}a^{2}\left(  t\right),
\end{equation}
the volume constraint becomes%
\begin{equation}
a^{3}\left(  t\right)  \sqrt{\widehat{\gamma}}=\frac{1}{3!}\epsilon
^{ijk}\epsilon_{abcd}h\left(  \sigma\right)  Y^{a}\partial_{i}Y^{b}%
\partial_{j}Y^{c}\partial_{k}Y^{d}\,. \label{a cube}%
\end{equation}
Equation (\ref{phi}) then reduces to
\begin{align}
\partial_{0}\left(  2\lambda a^{3}\right)   &  =\frac{1}{2}\partial_{0}\left(
\Lambda h\left(  \sigma\right)  \frac{1}{3!}\epsilon^{ijk}\epsilon_{abcd}%
Y^{a}\partial_{i}Y^{b}\partial_{j}Y^{c}\partial_{k}Y^{d}\right) \nonumber\\
&  =-2\partial_{0}\left(  \Lambda a^{3}\right)\,,
\end{align}
implying that
\begin{equation}
\lambda=-\Lambda+\frac{C}{2a^{3}}\,.%
\end{equation}
With $\widehat{\gamma}_{ij}=\delta_{ij},$ Eq.~(\ref{rho}) simplifies to
\begin{equation}
\frac{1}{a^{3}}\partial_{0}\left(  a^{3}\overset{.}{\sigma}\right)
=2\Lambda\frac{h^{\prime}\left(  \sigma\right)  }{h\left(  \sigma\right)  }\,.%
\end{equation}
The Einstein field equation (\ref{Einstein}) for the $0$-$0$ component is %
\begin{align}
G_{00}  &  =2\lambda+\frac{1}{2}\overset{.}{\sigma}^{2}+2\Lambda+U\left(
\sigma\right)  +T_{00}^{m}\nonumber\\
&  =\frac{1}{2}\overset{.}{\sigma}^{2}+\rho+U\left(  \sigma\right)  +\frac
{C}{2a^{3}}\,,%
\end{align}
where
\begin{equation}
\rho\left(  t\right)  =\rho_{0}\frac{a_{0}^{3}}{a^{3\left(  1+w\right)  }}\,.
\end{equation}
The equations for the $0i$-components are trivial, while, for the $ij$-components, they are%
\begin{align}
G_{ij}  &  =\left(  -\frac{1}{2}\overset{.}{\sigma}^{2}+2\Lambda+U\left(
\sigma\right)  \right)  g_{ij}+T_{ij}^{m}\\
&  =\left(  -\frac{1}{2}\overset{.}{\sigma}^{2}+2\Lambda-w\rho+U\left(
\sigma\right)  \right)  g_{ij}\,,%
\end{align}
where we have used the equation of state $p\left(t\right)  =w\rho\left(  t\right))\,,$ for some $w$ assumed to
be constant. For a Friedmann-Lema\^itre universe, the Einstein tensor $G_{\mu\nu}$ is given by
\begin{align}
G_{00}  &  =3\frac{\overset{.}{a}^{2}+k}{a^{2}}\,,\\
G_{ij}  &  =\left(  2\frac{\overset{..}{a}}{a}+\frac{\overset{.}{a}^{2}}%
{a^{2}}\right)  g_{ij}\,,%
\end{align}
and we obtain three differential equations for the two functions
$a\left(  t\right)  $and $\sigma\left(  t\right)$. Setting $k=0$ (and assuming that the 
self-interaction potential $U$ vanishes), these equations are given by
\begin{align}
3\frac{\overset{.}{a}^{2}}{a^{2}}  &  =\frac{1}{2}\overset{.}{\sigma}^{2}%
+\rho\left(  t\right) + \frac{C}{a^{3}}\,,%
\label{zero zero}\\
2\frac{\overset{..}{a}}{a}+\frac{\overset{.}{a}^{2}}{a^{2}}  &  =-\frac{1}%
{2}\overset{.}{\sigma}^{2}+2\Lambda-p\left(  t\right)\,,
\label{i j}\\
\overset{..}{\sigma}+3\frac{\overset{.}{a}}{a}\overset{.}{\sigma}  &
=2\Lambda\frac{h^{\prime}\left(  \sigma\right)  }{h\left(  \sigma\right)}\,, 
\label{rho equation}%
\end{align}
where $\rho\left(  t\right)  =\rho_{0}\frac{a_{0}^{3}}{a^{3\left(  1+w\right)
}}$ and $p\left(  t\right)  =w\rho\left(  t\right)$. It turns out that this system of equations 
cannot be solved exactly in presence of an arbitrary self-interaction potential $U\left(  \sigma\right)$,
except for special choices of $U$.
We can verify that the above three equations are not independent, by first multiplying 
equation(\ref{zero zero}) by $a^{3}\left(  t\right)$, then differentiating it with respect to time $t$, 
then dividing by $a^{2}\overset{.}{a}$, to obtain equation (\ref{i j}), after making
use of equation (\ref{rho}). We note that the the volume constraint,
\begin{equation}
a^{3}\left(  t\right)  \det\widehat{\gamma}=h\left(  \sigma\right)
\epsilon^{ijk}\epsilon_{abcd}Y^{a}\partial_{i}Y^{b}\partial_{j}Y^{c}%
\partial_{k}Y^{d}\,, \label{volumeconstraint}%
\end{equation}
implies that the ratio
\begin{equation}
\frac{a^{3}\left(  t\right)  }{h\left(  \sigma\right)  }=\frac{\epsilon
^{ijk}\epsilon_{abcd}Y^{a}\partial_{i}Y^{b}\partial_{j}Y^{c}\partial_{k}Y^{d}%
}{\det\widehat{\gamma}} \label{ratio}%
\end{equation}
has a right side that only depends on spatial coordinates and a left side that only
depends on time; hence they must be constant. This implies that
\begin{equation}
\partial_{0}\left(  \ln\frac{a^{3}\left(  t\right)  }{h\left(  \sigma\left(
t\right)  \right)  }\right)  =0\quad \Rightarrow\quad\overset{.}{\sigma}%
\frac{h^{\prime}\left(  \sigma\right)  }{h\left(  \sigma\right)  }%
=3\frac{\overset{.}{a}}{a}\,.%
\end{equation}
Assuming that visible matter and radiation can be neglected ($\rho_0 \equiv 0$) 
and that there is no mimetic dark matter ($C=0$), the equations to be solved are
\begin{align}
3\frac{\overset{.}{a}^{2}}{a^{2}}  &  =\frac{1}{2}\overset{.}{\sigma}^{2}\,,\label{first}\\
\overset{..}{\sigma}+3\frac{\overset{.}{a}}{a}\overset{.}{\sigma}  &
=2\Lambda\frac{h^{\prime}\left(  \sigma\right)  }{h\left(  \sigma\right)  }\,.
\label{second}%
\end{align}
Assuming that $h\left(  \sigma\right)  =e^{2\mu\,\sigma/f}$, we easily
see that these two equations are identical to equations (\ref{G00})
and (\ref{sigmaequation}). Thus this alternative formulation of the volume 
constraint, using the two maps $Y$ and $\widetilde{Y}$, of the manifold $M$ to two
three-spheres, is equivalent, locally, to the formulation using a closed three form, $\beta=dB$,
used in the bulk of the paper. The main difference between the discussion in this appendix and 
the earlier one lies in the use of the the ``winding fields'' $Y$ and $\widetilde{Y}$, instead of the
the two-form $B$, in the construction of the three-form $\beta$. The $Y$- and $\widetilde{Y}$ 
equations of motion are simple to solve, and we have shown that they lead to equivalent results
for$\sigma$ and the scale factor $a$.

\begin{center}
-----
\end{center}

\bigskip

\noindent
\href{mailto: chams@aub.edu.lb}{$^{*}$ chams@aub.edu.lb}
\\[0.3em]
\href{mailto: juerg@phys.ethz.ch}{$^{**}$juerg@phys.ethz.ch}.


\begin{thebibliography}{References}

\bibitem {CM}A.~H.~Chamseddine and V.~F.~Mukhanov, \textit{Mimetic Dark
Matter}, JHEP \textbf{11}, 135-140 (2013) [arXiv:1308.5410];\newline
A.~H.~Chamseddine and V.~F.~Mukhanov, \textit{Cosmology with Mimetic Matter},
JCAP \textbf{1406}, 017-031 (2014).

\bibitem {CCM}A.~H.~Chamseddine, A.~Connes and V.~F.~Mukhanov, \textit{Quanta
of Geometry: Noncommutative Aspects,} Phys.~Rev.~Letters \textbf{114}, 91302,
1-5 (2015), and references given there.

\bibitem {BF}R.~Brandenberger and J.~Fr\"ohlich, \textit{Dark Energy, Dark
Matter and Baryogenesis from a Model of a Complex Axion Field,} JCAP 04, 030
(2021); doi:10.1088/1475-7516/2021/04/030 [arXiv:2004.10025 [hep-th]].

\bibitem {Fr}J.~Fr\"ohlich, \textit{After the Dark Ages,} J. Phys. A: Math.
Theor. \textbf{55}, 421001 (2022); https://doi.org/10.1088/1751-8121/ac94aa

\bibitem {Wett}Chr.~Wetterich, \textit{Cosmology and the Fate of Dilatation
Symmetry}, Nucl. Phys. B \textbf{302}, 668 (1988);
doi:10.1016/0550-3213(88)90193-9 [arXiv:1711.03844].

\bibitem {Straumann}N.~Straumann, \textit{General Relativity and Relativistic
Astrophysics}, Texts and Monographs in Physics, Springer-Verlag, Berlin,
Heidelberg, New York 1984 (second printing 1991).

\bibitem {Stuck}E.~C.~G.~St\"uckelberg de Breidenbach, \textit{Die
Wechselwirkungskr\"afte in der Elektrodynamik und in der Feldtheorie der
Kernkr\"afte,} Helv.~Phys.~Acta \textbf{11}, 225 (1938).

\bibitem {Perl}S.~Perlmutter et al., \textit{Measurements of the Cosmological
Parameters} $\Omega$ \textit{and} $\lambda$ \textit{from the first Seven
Supernovae at} $z \geq0.35$, Astrophys.~J.~\textbf{483}, 565 - 581 (1997).

\bibitem {CFG}A.~Chamseddine, J.~Fr\"ohlich and O.~Grandjean, \textit{The
Gravitational Sector in the Connes-Lott Formulation of the Standard Model,} J.
Math. Phys. \textbf{36}, 6255-6275 (1995).

\bibitem{TW} M.~S.~Turner and L.~M.~Widrow, Phys. Rev. D \textbf{37}, 2743 (1988)

\bibitem{JS} M.~Joyce and M.~E.~Shaposhnikov, Phys. Rev. Lett. \textbf{79}, 1193-1196 (1997)
doi:10.1103/PhysRevLett.79.1193 [arXiv:astro-ph/9703005].

\bibitem{FP} J.~Fr\"ohlich and B.~Pedrini, in: ``Mathematical Physics 2000'', A.~Fokas, A.~Grigoryan, T.~Kibble and B.~Zegarlinski (eds.), 
Imperial College Press, London and Singapore 2000, pp. 9–47 [arXiv:hep-th/0002195v2]

\bibitem{Durrer} R.~Durrer and A.~Neronov, Astron. Astrophys. Rev. \textbf{21}, 62 (2013)
doi:10.1007/s00159-013-0062-7 [arXiv:1303.7121 [astro-ph.CO]].

\bibitem {Weinberg}S.~Weinberg, \textit{The Quantum Theory of Fields},
vols.~I-III, Cambridge University Press.

\bibitem{Einstein} A.~Einstein, \textit{Spielen Gravitationsfelder im Aufbau der materiellen Elementarteilchen 
eine wesentliche Rolle?} Sitzungsber. Preuss. Akad. Wiss. Berlin (Math. Phys.). 349-356 (1919).

\bibitem{Pauli} W.~Pauli, \textit{Theory of Relativity,} Dover, New York1981.

\bibitem{H-T} M.~Henneaux and C.~Teitelboim, \textit{The cosmological constant and general covariance,} 
Physics Letters B \textbf{222}, 195-199 (1989).

\end{thebibliography}
\end{document}